\documentclass[a4paper, 12pt]{article}  
\usepackage{geometry}
\usepackage{pdfpages}
\usepackage{lmodern}
\usepackage{graphicx} 
\usepackage{wrapfig} 
\usepackage{textcomp} 
\usepackage[utf8]{inputenc}
\usepackage[T1]{fontenc} 
\usepackage{float}
\usepackage{colortbl}
\newcommand{\hide}[1]{}
\DeclareSymbolFont{letters}{OML}{ztmcm}{m}{it}
\usepackage{color}
\usepackage{bm}
\renewcommand{\arraystretch}{1.2}
\usepackage{array}
\usepackage{listings}
\usepackage{adjustbox}
\usepackage[boxed]{algorithm2e}
\usepackage{subcaption}
\usepackage{multirow} 
\usepackage[OMLmathrm,OMLmathbf]{isomath}
\usepackage{booktabs} 
\usepackage{longtable}
\usepackage[hang,flushmargin]{footmisc}
\usepackage{pdflscape}
\usepackage{natbib}
\usepackage{makecell}
\usepackage[normalem]{ulem}
\usepackage{threeparttable} 
\usepackage{threeparttablex}
\usepackage{multibib}
\usepackage{makecell}
\usepackage{multicol}
\usepackage{tabularx}
\newcites{supp}{Supplementary Material References}

 \newcommand{\alert}[1]{\textcolor{red}{\bf{#1}}}

\usepackage{amsmath}
\usepackage{etoolbox} 
\usepackage{amsthm}
\usepackage{amsfonts}
\usepackage{amssymb}
\usepackage[section]{placeins}
\usepackage{diagbox} 
\usepackage[hidelinks]{hyperref}

\usepackage{fixmath} 
\usepackage{tikz}\usetikzlibrary{automata, arrows.meta, positioning}
\usepackage{footnote}
\makesavenoteenv{tabular}
\makesavenoteenv{table}
\usepackage{setspace}
\makeatletter
\renewcommand\@biblabel[1]{\textbf{#1.}} 
\usepackage{enumitem}
\newcommand{\blind}{1}
\usepackage{pifont}
 \newcommand{\comm}[1]{}
\setlist{nolistsep,leftmargin=*}

		\if1\blind
	{
	
\title{\textbf{Guardians of the Regime: Secret Police Formation in Autocracies}}
\author{
\vspace{-10pt}
Marius Mehrl$^1$, Mila Pfander$^2$, Theresa Winner$^3$, Cornelius Fritz$^4$\vspace{10pt}\\ 
\small School of Politics and International Studies, University of Leeds$^1$\vspace{-18pt}\\
\small
École Polytechnique Fédérale de Lausanne$^2$\vspace{-18pt}\\
\small Department of Statistics, LMU Munich$^3$\vspace{-18pt}\\
\small School for Computer Science and Statistics, Trinity College Dublin$^4$
}
\date{\normalsize\today}

	} \fi
	
	\if0\blind
	{
		\bigskip
		\bigskip
		\bigskip
		\title{\textbf{Guardians of the Regime: Secret Police Formation in Autocracies}}
		\medskip
	} \fi

\begin{document}

\vspace{-20pt}

\maketitle

\begin{abstract}
Autocrats use secret police to stay in power, as these organizations deter and suppress opposition to their rule. 
Existing research shows that secret police succeed at this but, surprisingly, also that they are not as ubiquitous in autocracies as one may assume, existing in fewer than half of autocratic country-years. 
We thus explore under which conditions secret police emerge in dictatorships. 
For this purpose, we develop a theoretical framework for potential predictors and apply statistical variable selection techniques to identify which of several candidate variables extracted from the literature on state security forces and authoritarian survival hold explanatory power.
Our results highlight that secret police are more likely to emerge when rulers face structural, regime-external threats, such as organised anti-system mobilisation and international rivals, or witness successful regime-internal contestation abroad that hints at similar threats at home. 
But additionally, we find that rulers must have sufficient material resources and personalised power to establish secret police. 
This research contributes to our understanding of autocrats' institutional choices and authoritarian politics.
\end{abstract}


\if1\blind
	{\noindent {\it Notes:} We thank Janina Beiser-McGrath, Olga Vlasova, the audience at the 2025 Political Studies Association Conference, Birmingham, and the anonymous reviewers and editor for their comments. 
	} \fi

\medskip

\pagebreak


\section{Introduction}
\label{sec:intro}

Secret police are a key instrument in the autocrat's quest to suppress opposition and remain in power. 
They are specialists in surveillance and preventive repression, as they are tasked with instilling fear, deterring dissident political mobilisation, and, where such mobilisation takes place, putting an end to it before it can escalate to threaten the ruler. 
Accordingly, existing work has shown that secret police presence is associated with reductions in citizens' individual and collective resistance \citep{hager2022does,HAGER_KRAKOWSKI_2024,choulis2024preventing}. 
While they can be broadly linked to increased physical repression \citep{mehrl2024secret}, there also is evidence that secret police succeed at deterring dissent. 
For instance in Eastern Germany, their presence is linked to reduced levels of political imprisonment \citep{Steinert_2023}, at least once they have established a reputation for tracking down and repressing opposition activity \citep{mehrl2024secret}.

Recent research has thus begun to elucidate how secret police, such as the Stasi, Pinochet's DINA/CNI, and Assad's Political Security Directorate, keep autocratic rulers in power.
Secret police can be defined as ``an internal security agency that uses violent policing
practices and intelligence operations against political opponents or dissidents while
being autonomous from the rest of the security apparatus'' \citep[p.995]{mehrl2024secret}. 
They thus differ from internal security institutions that report to the military, conduct operations combining violent policing and intelligence against criminal rather than political targets, target political opponents exclusively through intelligence work, or, like paramilitaries and militias, combine violence against such opponents with minimal intelligence activity\footnote{Additionally, militias differ from secret police due to their informal links to the state \citep{Carey_Colaresi_Mitchell_2015a}. 
} \citep[see][]{mehrl2024secret,Plate_Darvi_1982}. 

Earlier work has studied the principal-agent relationship between ruler and secret police, highlighting autocrats' incentives to limit the competence of their subordinates \citep{Egorov_Sonin_2011,Zakharov_2016,Dragu_Przeworski_2019,Thomson_2020}, but also the career pressures pushing subordinates to join secret police such as Argentina's Battalion 601 in the first place \citep{Scharpf_Gläßel_2020,Scharpf_Gläßel_2022}. More recently, research investigates how secret police navigate the trade-off between using the intelligence supplied by informants and the risk of their exposure \citep{Liu_Su_Wang_2026}.  
What may get lost across this body of work, however, is that secret police are not an automatic feature of autocratic governance. 
Indeed, global data shows that they exist in approximately a quarter of all military and personalist regime-years coded by \citet{Geddes_Wright_Frantz_2014}.
They are present in less than half of all party regime-years, have been highly prevalent only in the dictatorships of Europe and Central Asia, and, overall, existed in less than a third of all non-democracies over the period 1950--2018 \citep{choulis2024preventing,mehrl2024secret}. 
Given that secret police generally succeed in preventing dissent, this raises the question  -- when and where can we  expect secret police to be formed in autocracies?

Here, we answer this question. We develop a framework that distinguishes potential predictors of secret police formation along three dimensions --  willingness--opportunity, structure--shock, and regime internal--regime external -- and then identify candidate predictors from existing research on states' security force design and authoritarian survival. 
We apply the least absolute shrinkage and selection operator (LASSO, \citealp{tibshiraniRegressionShrinkageSelection1996}) to select the subset of these predictors that are associated with the onset of secret police organisations. 
This selection occurs by penalising the sum of the absolute values of the coefficients often leading to a number of non-zero entries.
Coefficients with little impact on the target variable are set to zero, effectively removing those predictors. 
This allows us to identify the variables that have the strongest influence on the formation of secret police.
Here, we apply a modified Lasso approach where stratification according to the outcomes in the cross-validation step is employed to ensure that the model is stable for binary outcomes with rare events, such as secret police formation. 
We find that secret police are most likely to be established when rulers not only perceive specific, structural rather than acute threats from outside of the regime, but also possess the necessary material resources and power base to act upon these threats.

This research advances the literature on authoritarian security institutions and autocracy more generally in several ways. 
First, we collect different theoretical arguments regarding the drivers of security institutional set-ups from existing studies, develop a theoretical framework to categorise them, and test them against each other in a principled fashion, finding that some prominent explanations contribute little to understanding secret police formation. 
Second, we provide the first study of when rulers choose to institute secret police organisations, thereby, third, providing a clear foundation for further research that may investigate the specific role of single predictors of secret police from a more theoretical and/or causal perspective. 
Finally, our research emphasises that what has regularly been considered a static feature of autocratic governance -- secret police -- is actually a variable to be explained, thus highlighting important variation in authoritarian institutional choices. 

\section{The Predictors of State Security Force Structure}
\label{sec:litrev}

Existing research has identified several predictors of state security force structure, that is, how governments decide to structure their security apparatus and what types of institutions they choose to invest in. This research has, in particular, studied the existence of counterbalancing paramilitaries and pro-government militias, but has also begun to investigate when governments recruit private military contractors and foreign legionnaires. As these types of security forces may share commonalities with secret police, but also important differences such as their main targets, emphasis on intelligence versus on violence, and clarity of links to the regime, their identified predictors provide a useful starting point to understand when rulers choose to establish a secret police.

Perhaps unsurprisingly, a first set of drivers of state security force structure very directly pertain to the risk of deposition the ruler is facing. Along these lines, several studies find that facing a higher risk of being deposed by regime elites via a coup d'etat, i.e., the level of internal threat, increases rulers' extent of coup-proofing, via the establishment of paramilitaries such as presidential guards \citep{Belkin_Schofer_2003,Belkin_Schofer_2005} or pro-government militias  \citep{Ash_2016,carey2016risk}, but also by hiring private military contractors \citep{Gentil-Fernandes_Morrison_Otto_2024}. But beyond coup risk, research also argues that rulers are susceptible to threats which are external to their regime, in particular mass-based mobilisation challenges in the form of protests or intrastate conflict \citep{carey2016risk,bohmelt2018auxiliary,Akins_2021,klosek2025one}, or even their country, pointing to the role of interstate rivalries and conflict \citep{Ash_2016, Akins_2021,Grasmeder_2021}. And \citet{Bohmelt_Ruggeri_Pilster_2017} focus again on the role of coup risk, but show that regimes also learn from external events as they react to the experience and behaviour of their ``peers'', that is, countries facing similar threats. 

These studies have also identified less direct, structural features of states that rulers can observe to learn about their level of threat and then make decisions on establishing specific security force types, but which may also constrain rulers in actually translating these decisions into action.  Along these lines, \citet{Pilster_Bohmelt_2012} show that democracies are less likely to coup-proof via paramilitaries, while \citet{Carey_Colaresi_Mitchell_2015a} highlight that pro-government militias are most likely to be formed in weakly democratic countries. But focusing specifically on autocracies, recent research also shows that modes of autocratic rule matter, with increased levels of personalist power concentration corresponding to an increased probability of having counterbalancing paramilitary forces \citep{Escriba_Bohmelt_Pilster_2020}, pro-government militias \citep{klosek2025one}, and foreign legionnaire recruitment \citep{mehrl2023dictator}. 

Finally, focusing on the constraints rulers face, \citet{bohmelt2018auxiliary} highlight that governments require more developed state capacity for specific security force configurations, arguing that mobilising, maintaining, and supervising paramilitaries requires more financial and administrative resources than linking up with militias would entail\footnote{Focusing on where states get their financial resources from, research on militias and private military contractors argues that reliance on democratic aid donors makes delegating violence to such ``external'' actors attractive  \citep{Carey_Colaresi_Mitchell_2015a,carey2016risk,Gentil-Fernandes_Morrison_Otto_2024}. However, secret police offer no such plausible deniability. Similarly, colonial inheritance is one source of paramilitaries, but secret police generally do not transfer from one regime to another \citep{Mehrl_Choulis_2021,mehrl2024secret}. We therefore do not consider these potential predictors below.}. 

\section{The (Potential) Predictors of Secret Police}
\label{sec:intuitions}

The insights generated by the literature on state security forces offer a starting point to identify factors that push rulers towards or away from establishing secret police. To facilitate this identification and the eventual systematic theorisation of factors, we distinguish them along three dimensions. 
First, willingness and opportunity \citep{most1984international}: is the variable expected to affect the ruler's motivation to establish a secret police, the constraints the ruler may have on this policy choice, or both? 
Second, structure versus disruptive events \citep{bermeo2016democratic,timoneda2023rush}: does a variable capture structural conditions which may affect the baseline probability of secret police formation, and gradually shift it over time, or a shock-like event indicating instantaneous changes in the dictator's willingness or opportunity for secret police formation? 
Third, regime versus external drivers \citep{BDM_2005,Roessler_2011}: is a factor located within the regime itself, for instance as a result of elite bargaining, power distribution, or regime emergence, or does it relate to the attributes and actions of regime outsiders?

These three dimensions offer a framework to think about, identify, and categorise the potential predictors of secret police formation, and more generally of policy shifts within the arena of authoritarian politics. 
Accordingly, the distinction between willingness and opportunity reflects what payoffs a leader expects, and what constraints they face, when forming a secret police. 
This phenomenon mirrors the distinction between militaries' disposition and ability to carry out a coup d’etat \citep{Finer_1962,Powell_2012}.
Here, willingness captures that secret police may assist dictators in navigating a challenging threat environment \citep{hager2022does,choulis2024preventing}, but also impose political costs and present threats to the dictator \citep{Egorov_Sonin_2011,Thomson_2020}. 
The opportunity dimension captures that dictators, even if willing to form a secret police, may be constrained to do so by their position vis-a-vis other elites \citep{Svolik_2012} but also a lack of material resources to train and maintain costly security forces \citep{bohmelt2018auxiliary}.
As such, the corresponding first dimension covers several dynamics ascribed to authoritarian decision-making, and captures whether a potential predictor of secret police affects the rulers' motivation or obstacles to form such an institution. 

In line with recent research on authoritarian politics \citep{timoneda2023rush}, the second and third dimension indicate whether it is a structural feature of or more of a shock-like shift to a given autocratic context, and whether it pertains to regime insiders or to actors and dynamics that are outside of it.
This framework, as illustrated in \autoref{tab:framework} with some of the variable discussed below, enables the identification and selection of variables to be included in the empirical models in a structured fashion, and facilitates model interpretation. 
Additionally, it provides a bridge point for the study of potential further predictors of secret police emergence, allowing for them to be straightforwardly connected to the insights of the models we present here.

\begin{table}[h!]
\scriptsize
\caption{(Potential) predictors of Secret Police along three dimensions.}
\label{tab:framework}
\centering
\begin{subtable}[t]{0.48\textwidth}
\centering
\begin{tabular}{cp{2.5cm}p{2.5cm}}
\toprule
 & Structure & Shock \\ \cline{2-3}
Willingness & Regime \& leader duration & Coup attempt \\
Opportunity & Personalism; State capacity & Coup attempt \\
\bottomrule
\end{tabular}
\caption{Regime-internal predictors}
\end{subtable}
\hfill
\begin{subtable}[t]{0.48\textwidth}
\centering
\begin{tabular}{cp{2.5cm}p{2.5cm}}
\toprule
 & Structure & Shock \\ \cline{2-3}
Willingness & Excluded ethnic population; substitute security forces & Protests; Neighbour coups \& protests \\
Opportunity & Wealth; Oil \& gas production & Economic Growth \\
\bottomrule
\end{tabular}
\caption{Regime-external predictors}
\end{subtable}
\end{table}

That being said, our goal here is to provide a first indication of which of these factors, and, thus, the theoretical dynamics discussed below, are associated with the establishment of secret police institutions. We thus do not develop a specific, fully formulated theory for each of them. Instead, we lay the groundwork for such theory development via the framework introduced here and exploring which of the several potential theoretical dynamics actually appear worthy of investigation\footnote{We also opt against developing theoretical hypotheses as testing them together would likely prove impossible \citep{Keele_Stevenson_Elwert_2020}.}.  

In line with earlier research on security force structure, we include several indicators of events directly threatening the ruler in our model.
For instance, coup attempts are regime-internal shocks that may shift the ruler's willingness and opportunity to form a secret police, while other threats at home, such as protests, civil and interstate conflict, or those occurring in their neighbourhood represent external shocks. In contrast, autocratic regime type and financial and administrative state capacity represent structural factors linked to the regime which should mainly affect the dictator's opportunity to institute secret police, but leave their willingness to do so comparatively unaltered. These factors are placed accordingly into \autoref{tab:framework} and we list their operationalization into variables to be included in our models in \autoref{tab:selection}.  



Beyond these potential predictors taken from existing research on state security forces, we next turn to the broader literature on authoritarian politics and survival to identify further potential factors within and outside of the regime, structural or shock-like, that may affect dictator's opportunity and/or willingness to create secret police. First, we include  several structural covariates that capture societal risk factors associated with anti-government mobilisation, which should increase the ruler's risk perception and, accordingly, willingness to invest into a secret police. 
Therefore, we incorporate covariates regarding the population size, the population share of politically included and excluded ethnic groups, as well as what percentage of the population lives in urban settings. 
These variables have been prominently linked to regime threats and survival strategies \citep[see e.g.][]{Roessler_2011,Chenoweth_Ulfelder_2017}. Along the same lines, but capturing closer structural precursors of political mobilisation, we also include measures of civil society independence, participation level, and the presence of anti-government civil society organisations. And as rulers may well expect mass mobilisation, and thus the need for secret police, to increase disproportionately around elections, we include a dummy variable for election years. 

Similarly, we add further structural characteristics of the regime that capture how established and cohesive the dictator's rule is, and hence should affect their perceived need and according willingness to create secret police. Along these lines, we include variables counting the time the current leader and regime have been in power \citep{Svolik_2012}. We include a measure capturing whether a regime uses ideological legitimisation claims, following the notion that such regimes may have reduced needs for security institutions as their stability arises from shared goals \citep{Escriba_Bohmelt_Pilster_2020}\footnote{Note that alternatively, ideological legitimisation claims may also trigger a stronger need to protect their core ideological tenets among their citizens, and thus increase regime willingness to form secret police.}. And given that the ruler's expected payoffs from, and thus willingness to invest in, another repressive agent likely depend on their existing efforts at tackling threats, we include covariates that gauge the level of repression, both generally and targeted specifically at civil society organisations.

In light of this, we also include as a structural regime feature the existence of other security institutions with relatively similar tasks as secret police -- counterweight paramilitaries, affiliated forces, and the overall degree of counterbalancing. Given that all of these actors can, at least to some degree, be used to counter threats emanating from within the regime as well as other domestic sources of danger, they may substitute for each other, with the presence of one reducing the ruler's need to create another. For the same reason, we include a variable capturing clientelistic rule, which offers an alternative mode of generating information and control, thus reducing the need for a formal secret police. 

In theory, a ruler may also choose to create endless parallel security institutions to counterbalance each other, but forming even one such institution requires extensive resources (\citealp{bohmelt2018auxiliary}; \citealp[p.157]{Geddes_Wright_Frantz_2018}). As such, we include several structural features of the regime's resource environment, including the country's general economic performance, defence spending, as well as oil and gas production as a source of disposable income \citep{fails2020oil}. These structural characteristics should affect dictator's opportunities to fund the formation of new secret police institutions, as should shock-like shifts in economic growth.

\autoref{tab:selection} lists all covariates included in our models, specifying how we operationalise the variables discussed above. Note that in several cases, we include various operationalizations and leave selecting among them to the model. We report data sources for all included variables in the supplementary materials.

\hide{
\begin{table}[h]
    \caption{Variable Selection by LASSO and Stepwise Methods. Model 1 represents the LASSO model with a logit link, while Model 2 corresponds to the LASSO model with a cloglog link. Model 3 corresponds to the Stepwise model with a logit link, and Model 4 to the Stepwise model with a cloglog link.}
    \label{tab:selection}
    \centering
    \begin{adjustbox}{max width=\textwidth}
    \begin{tabular}{llcc}
        \toprule
        & Variable & Selected in Model(s) \\
        \midrule
  Intrastate Conflict \citep[UCDP;][]{davies2024organized}      & intrastate         &  - \\
   Democracy Score \citep[Polity5;][]{Marshall_Gurr_2020}    & polity2            &  -\\
   Coup Attempt \citep{Powell_Thyne_2011}     & attempt            & - \\
  Human Rights \citep[Latent score;][]{Fariss_2019}      & theta\_mean        & 1, 2 \\
   State Capacity \citep[Latent score;][]{Hanson_Sigman_2021}     & Capacity           & - \\
    Bureaucratic Capacity \citep[V-DEM;][]{Coppedge_Gerring_Knutsen_Lindberg_Teorell_etal._2023}   & v2clrspct          & - \\
     Fiscal Capacity \citep[V-DEM;][]{Coppedge_Gerring_Knutsen_Lindberg_Teorell_etal._2023}   & v2stfisccap        & - \\
  Territorial Control \citep[V-DEM;][]{Coppedge_Gerring_Knutsen_Lindberg_Teorell_etal._2023}      & v2terr             & - \\
 CSO entry and exit \citep[V-DEM;][]{Coppedge_Gerring_Knutsen_Lindberg_Teorell_etal._2023}       & v2cseeorgs         & - \\
       CSO repression \citep[V-DEM;][]{Coppedge_Gerring_Knutsen_Lindberg_Teorell_etal._2023} & v2csreprss         & 1, 2, 3, 4  \\
 CSO participatory environment \citep[V-DEM;][]{Coppedge_Gerring_Knutsen_Lindberg_Teorell_etal._2023}      & v2csprtcpt         &  1 \\
     CSO anti-system movements \citep[V-DEM;][]{Coppedge_Gerring_Knutsen_Lindberg_Teorell_etal._2023}  & v2csantimv         & 1, 2, 3, 4 \\
 CSO Strength  \citep[V-DEM;][]{Coppedge_Gerring_Knutsen_Lindberg_Teorell_etal._2023}     & v2csstruc\_1       &  -  \\
 Regime change \citep[CHISOLS;][]{Mattes_Leeds_Matsumura_2016}     & solschdum          &  - \\
    Urban Population \citep[\%;][]{WorldBank_2021}    & urbanpop           & - \\
   Economic Growth \citep{WorldBank_2021}    & l12gr              & - \\
   Personalisation \citep[Latent score;][]{Geddes_Wright_Frantz_2018}     & xpers              & 3, 4 \\
   Ethnically excluded population \citep[\%;][]{Vogt_Bormann_Rüegger_Cederman_Hunziker_Girardin_2015}    & lexclpop           &  - \\
Counterbalancing \citep{Pilster_Bohmelt_2011}       & effectivenumber    & - \\
Military Expenditures \citep[Latent score, logged;][]{Barnum_Fariss_Markowitz_Morales_2024}        & mean\_log          & 1, 2, 3, 4 \\
 Oil production \citep{Ross2015_oildata}       & oil\_value\_2000   &  1, 2\\
 Gas production \citep{Ross2015_oildata}       & gas\_value\_2000   &  1, 2, 3, 4\\
 Election year \citep[NELDA;][]{Hyde_Marinov_2012}        & election           & - \\
 Leader Duration \citep[CHISOLS;][]{Mattes_Leeds_Matsumura_2016}       & leader\_duration   & 1  \\
  Regime Duration \citep[CHISOLS;][]{Mattes_Leeds_Matsumura_2016}      & regime\_duration   & 1, 2, 3, 4\\
 Failed Coups in Region \citep{Powell_Thyne_2011} & unsucc\_coups\_reg & -  \\
  Successful Coups in Region \citep{Powell_Thyne_2011}   & succ\_coups\_reg   & - \\
 International Rivalry: Dummy \citep{Thompson_Dreyer_2011}       & rivalry\_dummy     & 1, 2, 3, 4 \\
  International Rivalry: Count  \citep{Thompson_Dreyer_2011}    & rivalry\_count     & 1, 2, 3, 4 \\
   Militarised Interstate Dispute: Dummy \citep{Palmer_McManus_D’Orazio_Kenwick_Karstens_Bloch_Dietrich_Kahn_Ritter_Soules_2022}      & mid\_dummy         &  - \\
  Militarised Interstate Dispute: Count \citep{Palmer_McManus_D’Orazio_Kenwick_Karstens_Bloch_Dietrich_Kahn_Ritter_Soules_2022}     & mid\_count         & 1, 3, 4\\
 Protest \citep[Latent score;][]{Chenoweth_D’Orazio_Wright_2014}       & protest            & 1, 2, 3, 4 \\
   Neighbour Protest \citep[Latent score;][]{Chenoweth_D’Orazio_Wright_2014}      & nbr\_protest       & -   \\
  GDP p.c. \citep[Logged;][]{WorldBank_2021}     & ln\_gdp\_pc        & 1, 2  \\
Population \citep[Logged;][]{WorldBank_2021}       & ln\_pop            &  - \\
        \bottomrule
    \end{tabular}
    \end{adjustbox}
\end{table}

}

\hide{
\begin{table}[t]
    \caption{Variable Selection by LASSO and Stepwise Methods: We performed LASSO selection using a logit link (Model 1) and cloglog link (Model 2) and Stepwise selection using a logit link (Model 3) and cloglog link (Model 4).}
    \label{tab:selection}
    \centering
    \begin{tabular}{llc}
        \toprule
         Variable & Selected in Model(s) \\
        \midrule
  Intrastate Conflict \citep[UCDP;][]{davies2024organized}           &  - \\
   Democracy Score \citep[Polity5;][]{Marshall_Gurr_2020}         &  -\\
   Coup Attempt \citep[Time since event;][]{Powell_Thyne_2011}\footnote{\alert{Still to be included in this manner, same operationalization will also be used for intrastate and interstate conflict.}}            & - \\
  Human Rights \citep[Latent score;][]{Fariss_2019}             & 1, 2 \\
   State Capacity \citep[Latent score;][]{Hanson_Sigman_2021}        & - \\
    Bureaucratic Capacity \citep[V-DEM;][]{Coppedge_Gerring_Knutsen_Lindberg_Teorell_etal._2023}       & - \\
     Fiscal Capacity \citep[V-DEM;][]{Coppedge_Gerring_Knutsen_Lindberg_Teorell_etal._2023}     & - \\
  Territorial Control \citep[V-DEM;][]{Coppedge_Gerring_Knutsen_Lindberg_Teorell_etal._2023}                  & - \\
 CSO entry and exit \citep[V-DEM;][]{Coppedge_Gerring_Knutsen_Lindberg_Teorell_etal._2023}            & - \\
       CSO repression \citep[V-DEM;][]{Coppedge_Gerring_Knutsen_Lindberg_Teorell_etal._2023}     & 1, 2, 3, 4  \\
 CSO participatory environment \citep[V-DEM;][]{Coppedge_Gerring_Knutsen_Lindberg_Teorell_etal._2023}           &  1, 2 \\
     CSO anti-system movements \citep[V-DEM;][]{Coppedge_Gerring_Knutsen_Lindberg_Teorell_etal._2023}          & 1, 2, 3, 4 \\
 CSO Strength  \citep[V-DEM;][]{Coppedge_Gerring_Knutsen_Lindberg_Teorell_etal._2023}           &  -  \\
 Regime change \citep[CHISOLS;][]{Mattes_Leeds_Matsumura_2016}     &  - \\
    Urban Population \citep[\%;][]{WorldBank_2021}            & 4 \\
   Economic Growth \citep{WorldBank_2021}             & - \\
   Personalisation \citep[Latent score;][]{Geddes_Wright_Frantz_2018}           & 1, 2, 3, 4 \\
   Ethnically excluded population \citep[\%;][]{Vogt_Bormann_Rüegger_Cederman_Hunziker_Girardin_2015}       &  - \\
Counterbalancing \citep{Pilster_Bohmelt_2011}      & - \\
Military Expenditures \citep[Latent score, logged;][]{Barnum_Fariss_Markowitz_Morales_2024}            & 1, 2, 3, 4 \\
 Oil production \citep[Logged financial value;][]{Ross2015_oildata}       & - \\
 Gas production \citep[Logged financial value;][]{Ross2015_oildata}         &  3, 4\\
 Election year \citep[NELDA;][]{Hyde_Marinov_2012}               & - \\
 Leader Duration \citep[CHISOLS;][]{Mattes_Leeds_Matsumura_2016}        & 1, 2  \\
  Regime Duration \citep[CHISOLS;][]{Mattes_Leeds_Matsumura_2016}         & 1, 2, 3, 4\\
 Failed Coups in Region \citep{Powell_Thyne_2011}  & -  \\
  Successful Coups in Region \citep{Powell_Thyne_2011}     & - \\
 International Rivalry: Dummy \citep{Thompson_Dreyer_2011}           & 1, 2, 3, 4 \\
  International Rivalry: Count  \citep{Thompson_Dreyer_2011}      & 1, 2, 3, 4 \\
   Militarised Interstate Dispute: Dummy \citep{Palmer_McManus_D’Orazio_Kenwick_Karstens_Bloch_Dietrich_Kahn_Ritter_Soules_2022}           &  - \\
  Militarised Interstate Dispute: Count \citep{Palmer_McManus_D’Orazio_Kenwick_Karstens_Bloch_Dietrich_Kahn_Ritter_Soules_2022}           & 1, 2, 3, 4\\
 Protest \citep[Latent score;][]{Chenoweth_D’Orazio_Wright_2014}            & 1, 2, 3, 4 \\
   Neighbour Protest \citep[Latent score;][]{Chenoweth_D’Orazio_Wright_2014}           & -   \\
  GDP p.c. \citep[Logged;][]{WorldBank_2021}            & 1, 2, 3  \\
Population \citep[Logged;][]{WorldBank_2021}                  &  - \\
        \bottomrule
    \end{tabular}
\end{table}
}

\begin{table}[t]
  \begin{threeparttable}
\scriptsize
    \caption{Variable Selection by LASSO and Stepwise Methods. LASSO selection was performed using a logit link (Model 1) and cloglog link (Model 2); Stepwise selection was performed using a logit link (Model 3) and cloglog link (Model 4).}    \label{tab:selection}
    \centering
\begin{tabular}{  m{5.6cm}  m{1.4cm}  m{5.6cm} m{1.4cm}  }         \toprule
        \textbf{Variable} & \textbf{Selected in Model} & \textbf{Variable} & \textbf{Selected in Model} \\
        \midrule
  Intrastate conflict: Dummy &  - & Clientelism & - \\
  Intrastate conflict: Years since &  - &   Ethnically excluded population (\%) &  - \\
  Neighborhood intrastate conflicts (count) & - & Counterweight Forces & -\\
   Democracy score  &  - & Counterbalancing     & - \\
   Coup attempt: Dummy  & - &  Military expenditures (Latent score, \emph{ln})           & 1, 2, 3, 4 \\
   Coup attempt: Years since    & - &  Oil production (Financial value, \emph{ln})    & - \\
  Human rights (Latent score) & 1, 2 &  Gas production (Financial value, \emph{ln})         &  3, 4\\
   State capacity (Latent score)       & - &  Election year             & - \\
    Bureaucratic capacity      & - &  Leader duration        & 1, 2 \\
     Fiscal capacity      & - &   Regime duration        & 1, 2, 3, 4\\
  Territorial control                  & - &  Neighborhood unsuccessful coups (count)  & -  \\
 CSO entry and exit   & - & Neighborhood successful coups (count) & 1, 2, 3, 4 \\
       CSO repression   & 1, 2, 3, 4  &  International rivalry: Dummy           & 3, 4 \\
 CSO participatory environment    &  1, 2 &   International rivalry: Count      & 1, 2 \\
     CSO anti-system movements       & 1, 2, 3, 4 &   MID: Dummy  &  - \\
 CSO strength    &  -  &    MID: Years since  &  - \\
 Regime change  &  - &   MID: Count  & 1, 2, 3, 4\\
    Urban population (\%) & - &  Protest (Latent score)            & 1, 2, 3, 4 \\
   Economic growth             & - & Neighborhood protest (average score) & 1, 2   \\
   Personalisation (Latent score)  &  1, 2, 3, 4 &   GDP p.c. (\emph{ln})           & 1, 2, 3, 4  \\
   Ideological legitimisation & 3, 4 & Population (\emph{ln})                  &  - \\
   Affiliated forces & - & Neighborhood: Secret Police (count) & -\\
        \bottomrule
    \end{tabular}
  \end{threeparttable}
\end{table}

\section{Research Design}

Our analyses are based on a dataset covering 120 autocratic countries over the period 1951--2018 from \citet*{choulis2024preventing}. 
The target variable $y_{i,t}$ is a binary indicator of secret police onset in country $i$ and year $t$, taking the value 1 if a secret police is established and 0 otherwise, where secret police are coded in line with the definition introduced above along five criteria capturing their targeting of political opponents, operational independence and secrecy, specialisation in intelligence operations, and regular use of violent policing practices \citep{choulis2024preventing,mehrl2024secret}. 
Following \citet{McGrath_2015}, we set $y_{i,t}$ to missing if a secret police exists but was established before. 
If the transition is interrupted by a gap in the dataset due to the suspension of autocracy, this is not counted as a formation. 
As a result, we disregard any formations at time point $t$, if, e.g., there was no autocracy or observation at $t-1$.
According to this scheme, we observe 31 secret police formations in 29 countries, with Iran and Cuba each experiencing two formations.
The dataset contains $\mbox{3,759}$ observations in total.
We include all independent variables listed in \autoref{tab:selection} lagged by one year to account for the time between changes in the covariates and their potential effect on secret police formation. \hide{During data preparation, covariates with a proportion of over 40 \% of missing values are removed\footnote{The corresponding covariates are not included in Table \ref{tab:selection}}.} 
Missing values are imputed using bootstrap-based methods with the \texttt{Amelia} package \citep{honaker2011}\footnote{For this imputation step, we require that the missing values are missing at random (MAR), meaning that the probability of missingness depends only on observed data and not on the missing values themselves. 
This is a plausible assumption for our data.}. 
Thus, we make use of multiple imputation in an iterative algorithm that, first, imputes missing values and, second, maximises the likelihood of the complete data. 
To account for uncertainty introduced the imputation step, we adopt a bootstrapping approach.
For robustness, we impute the dataset five times, apply our methodology to each imputed version, and then average the resulting models to obtain more stable estimates. As we are interested in when and where secret police are formed, we make use of both variation across countries and across time.

The employed methodology follows a two-step approach: first, we identify the most relevant covariates via statistical variable selection, and second, we examine the magnitude, direction, and significance of their effects. The second step simply amounts to estimating logistic regression models, as we fit a logistic model on the selected variables for each dataset and pool the results by averaging. However, the first step requires a more detailed explanation.

Since the sample size of our data is relatively small and our aim is to identify a parsimonious set of covariates  from \autoref{tab:selection}, we first fit a Lasso model to the data where the absolute value of all parameters is penalised with a parameter $\lambda$ \citep{tibshiraniRegressionShrinkageSelection1996}.  
This penalisation will set the parameters of some coefficients to zero and thus detects which variables are associated with the establishment of secret police organisations.
Since the dependent variable is binary, we employ a logit link function. 
In line with standard practice \citep{tibshiraniRegressionShrinkageSelection1996,hastie2009elements,friedman2010regularization}, the penalisation parameter $\lambda$ was chosen by minimising the deviance through three-fold cross validation.  
Given that there are only 31 observed formations, we modify the standard Lasso by stratifying each cross-validation dataset to ensure that every fold contains approximately the same number of secret police establishments. This ensures a balanced distribution across the folds, and makes the Lasso model also applicable to binary outcomes where one value is rare.
The stratified Lasso model is applied to each of the five imputed datasets detailed in the previous paragraph. We keep only those variables that appear in at least three out of the five models fitted on the different imputed datasets.  
Finally, we fit an unconstrained logistic model with the selected variables for each dataset and pool the results via Rubin's (\citeyear{rubin1987multiple}) rule.

To further validate our findings, we repeat the analysis using an alternative stepwise selection approach based on Akaike Information Criterion (AIC) instead of Lasso. 
This procedure begins with an empty model and iteratively adds the variable minimising the AIC at each step until no further improvement is possible. 
While straightforward and easy to interpret, this approach makes local, step-by-step decisions that may not yield the overall optimal set of predictors.
Moreover, results are often sensitive to the order in which variables are introduced \citep{hastie2020best}. 
As such, we prioritise the Lasso and use the stepwise selection approach as a robustness check, ensuring that our results are not specific to one particular variable selection method.
Reassuringly, \autoref{tab:selection} shows that the variables selected via the two approaches are closely aligned.

As a further robustness check, we replace the logit-link with a complementary log-log (cloglog) link function. 
Unlike the symmetric logit transformation, the cloglog-link is asymmetric and commonly used for modelling binary outcomes where one class is rare \citep{tutz2016modeling}, as is the case with our dependent variable. 
The results using  the cloglog-link remain similar to those from models fitted with the logit-link.




\section{Empirical Results}

The first set of results, that is, which variables were selected by which approach (Lasso with logit link, Lasso with cloglog link, stepwise selection with logit link, and stepwise selection with cloglog link models), is provided in Table \ref{tab:selection}. 
We observe that several of the potential predictor variables were not selected by all four selection approaches. 
None of the several measures of state capacity was selected in a single model, suggesting that this institutional dimension is not associated with autocrats' decision to set up secret police. 
There is also no evidence that the occurrence or a history of coup attempts or civil war influence whether secret police are created.  
Further, existing alternative approaches to managing threats to the regime, such as counterbalancing or clientelism, were left out of all models, as were demographic variables. 
Finally, intrastate conflict and protests in the neighbourhood do not appear to play a role for creating secret police. At the same time, Table \ref{tab:selection} exhibits largely compatible results between the four variable selection approaches, indicating that overall results do not depend on a particular selection approach or link function.

\begin{figure}
    \centering
    \includegraphics[width=\linewidth]{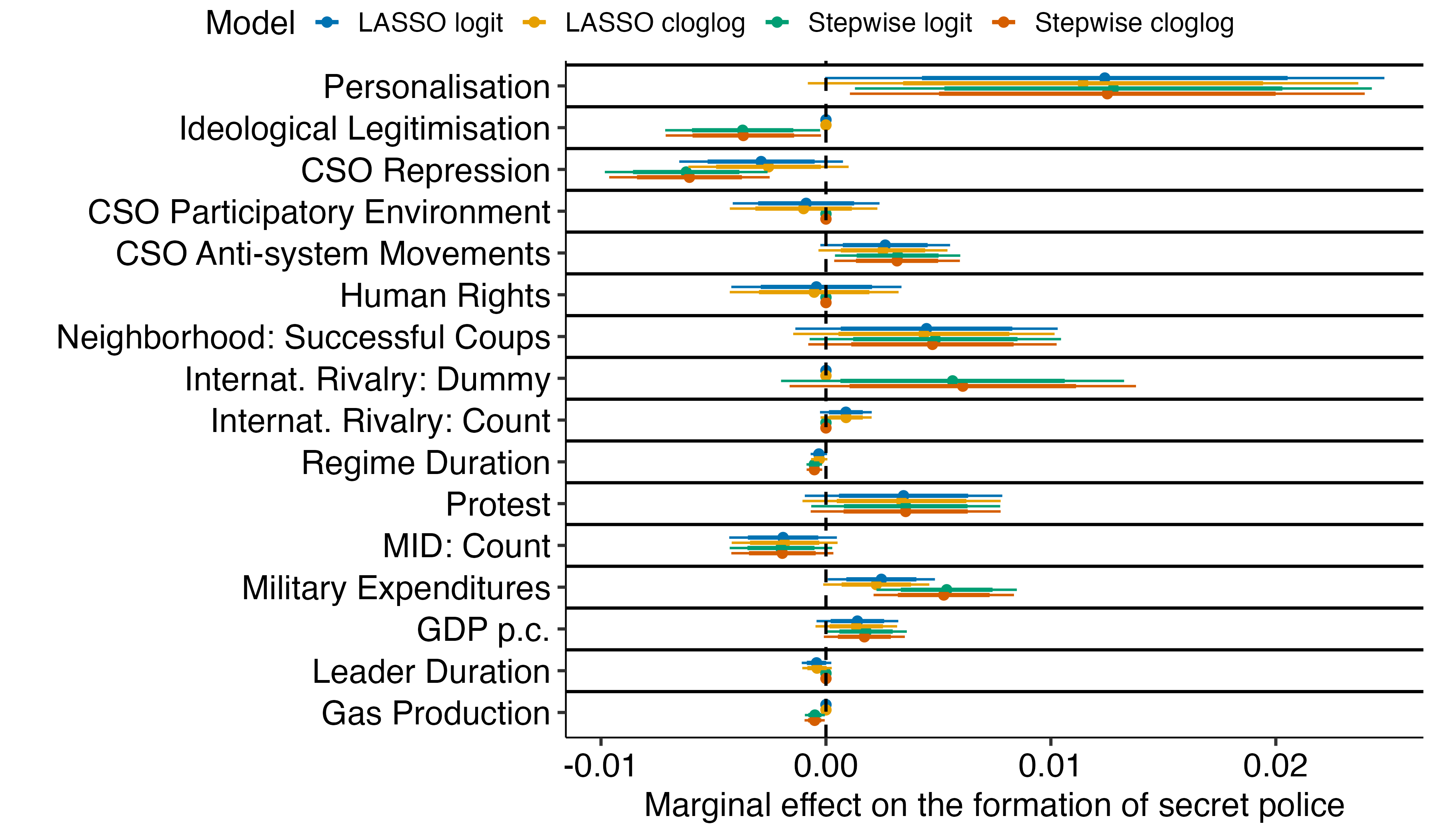}
    \caption{Predictors of Secret Police: Marginal effect estimates with 90\% and 95\% confidence intervals (thick and thin whiskers, respectively).
    Each variable is associated with four coefficient estimates, corresponding to LASSO and Stepwise selection methods applied with logit and cloglog link functions. 
    Coefficients for unselected variables plotted as zero to maintain comparability across approaches.
     }
    \label{fig:marginal_effects}
\end{figure}

Figure \ref{fig:marginal_effects} displays the marginal effects on the probability of secret police formation for all variables selected by at least one of the employed methods\footnote{See the Supplementary Materials for results tables.}. 
We restrict the plot to variables identified by at least one of the four approaches detailed in Table \ref{tab:selection}. 
Figure \ref{fig:marginal_effects} indicates that none of the detected effect sizes is large -- however, this is unsurprising given how rare secret police formations are.

Looking at specific effect estimates, Figure \ref{fig:marginal_effects} supports several of the intuitions developed in Section \ref{sec:intuitions} about secret police formation. 
First, the autocrat's external threat environment appears to be important: secret police are more likely to be established if civil society organisations challenging the system of government exist and are stronger, protests are more rife, and if the country is operating within an international rivalry. Interestingly, this positive relationship between threat and secret police formation appears limited to more latent, structural threats which have not yet fully escalated. 
This interpretation is supported by the finding that variables capturing more shock-like events, such as intrastate conflict, coup attempts, and the presence of interstate dispute, remain unselected, while the count of interstate disputes is found to have a negative relationship with secret police establishment. 
The exception to this pattern is the positive coefficient estimate recovered for successful coups in the neighbourhood, which, though shock-like, indicates that autocrats observe such events, update their beliefs about the threats they are facing themselves, and move to adjust their security apparatus accordingly \citep{ben2025following}. 
In this sense, shock-like events elsewhere change the dictator's belief about their own structural sources of danger. 
That successful coups in the neighbourhood are found to increase the probability of secret police establishment, while none of the domestic indicators of coup risk are selected, may further imply that dictators form secret police to stop military threats to their rule from emerging, rather than to tackle them once such threats are realised in the form of, for instance, politicised officer cliques within the military or rumors of coup plans. 
And more broadly, these conclusions suggest that secret police are formed to pre-empt and be prepared for escalation when the structural threat environment makes it appear likely, as opposed to when escalation actually takes place. In other words, secret police may be created to delay or even stop an endgame from occurring, not to help autocrats survive the endgame when it is already occurring.  

Indeed, when faced with a potential endgame, creating new coercive institutions may only serve to further weaken the ruler's position vis-a-vis the ruling elite instead of strengthening it \citep[see][]{Svolik_2012}, or be unattractive as the ruler actually enjoys increased elite loyalty for the time being \citep{Mcmahon_Slantchev_2015}. 
This first set of results thus highlights the importance of the regime-external structural threat environment, which alters autocrats' willingness \emph{and} opportunity for policy changes, in predicting secret police formation. However, these threats appear to have to be realised to some extent, such as in the form of anti-system organisations or protests, as our results indicate structural predictors of threats, including recent intrastate conflict experience, civil society strength, or excluded ethnic population share, to be of negligible importance.  

Regarding regime-internal structural characteristics, we find that secret police formation is less likely if levels of civil society repression are already high, and when the regime is in place for longer. As discussed above, both of these characteristics should reduce rulers' perceived need, and hence their willingness, to invest in additional regime survival policies. In such cases, a secret police will thus either already exist or is not necessary any more as the regime is protected via other means. Along these lines, there is also some evidence, though found only select models, that autocrats are less likely to establish secret police if the specific leader has been in power for longer or they can rely on ideological claims to legitimise their rule. There is thus evidence that several regime-internal structural features, which we propose to reduce the need for and thus willingness to invest in secret police, to be associated with a lower probability of secret police formation. Interestingly, however, this finding does not extend to the presence of alternative security actors, such as paramilitaries, or sources of information and control, such as clientelism, as these remain unselected across models.       
 
Moreover, while none of the state capacity indicators were selected, the results in Figure \ref{fig:marginal_effects} offer support to the idea that the regime's material resources act as a structural predictor of its ability to fund and maintain, and hence establish, a secret police. 
Specifically, military spending is positively associated with secret police formation, suggesting that as autocracies spend more on their security apparatus, they also have more opportunity to form a secret police. GDP per capita equally exhibits a positive effect estimate, suggesting a similar interpretation. Interestingly, oil and gas production remained generally unselected, indicating that it is not specifically disposable income which matters for whether secret police are established, but a regime's general material resources.
Finally, our models indicate that personalism, another structural characteristic of the regime which above is argued mainly affect their ruler's ability to create secret police, matters: rulers who have concentrated more power on themselves are more likely to establish a secret police, with increasingly marginalised elites being less able to resist this move.

In the supplementary materials, we detail data sources and present a results table for the models underlying Figure \ref{fig:marginal_effects}. We apply these models to predict counterweight paramilitaries instead of secret police, and find that while many of the statistically significant variables are they same, their effect directions are flipped. This may indicate that rulers form these distinct security actors when facing different threats and conditions. And we estimate additional models where we additionally allow continuous covariates to enter the model as first differences and construct indicators of the time since their last realisation for all binary covariates.   

Taken together, our empirical results suggest that autocrats establish secret police when facing a structural external threat environment, indicating opposition from their own citizens or international rivals. In contrast, other domestic threats, including coups and civil war, appear not to matter, suggesting that secret police may be created to pre-empt, rather than tackle actualised threats, and to mainly target threats from outside the regime rather than elites. Additionally, our results highlight that autocrats are more likely to establish secret police when their structural resource and power base allows them to do so, while the probability of secret police formation decreases if regime-internal structure indicates that such an additional security organisation is less needed.

\section{Conclusion}
Secret police, such as the Soviet Cheka, NKVD, and KGB, Franco's Political-Social Brigade, and Saddam's Mukhabarat, play a key role in ensuring authoritarian rulers' survival. 
But despite this and the prominence of many such organisations, secret police exist in less than a third of non-democratic country-years where data is available to us.
Motivated by this apparent disconnect between the usefulness and commonness of secret police, we investigate when secret police are established in autocracies. 
This paper develops a theoretical framework to identify and interpret predictors in a structured fashion, extends existing statistical variable selection techniques to better handle rare binary outcomes such as secret police formation, and applies the resulting method to assess which of the several potential predictors are actually meaningfully associated with the establishment of secret police.   

Our results highlight that secret police are most likely to be formed when rulers face particular, structural rather than shock-like threats external to their regime, but also have the opportunity to do so. 
As such, our results point to why, as highlighted above, secret police might be so surprisingly rare in autocracies: 
Rulers have to face a fairly specific combination of threats and their own resources to be both able and willing to create a secret police. As such, secret police formation depends not only on rulers' access to sufficient material resources to fund such an organisation, but also their exposure to threats which are regime-external and sufficiently important, but also not too acute. Otherwise, rulers appear to opt against creating secret police, potentially due to the very real costs these institutions can impose on rulers \citep{Egorov_Sonin_2011,Thomson_2020}. Connected to this, our data also suggest that an additional reason for secret police formation being so rare is that secret police tend to stick around. In other words, rulers have a hard time dissolving one and then establishing another secret police, further re-iterating that choosing to create such a powerful security agency is not an automatic decision for autocrats.

For future research, we offer both tools and directions to further advance our understanding of secret police, state security force structure, and autocrats' strategies of survival. We provide a flexible theoretical framework to systematise and interpret potential predictor variables, thus allowing future work to straightforwardly add new covariates to those studied here. The stratified Lasso method developed for this research offers a directly applicable approach to variable selection for predicting secret police formation, but also a range of other relevant rare outcomes. And together with the data and code for this paper, these contributions will serve as a starting point for future studies that want to expand the range of considered predictor variables, investigate alternative dependent variables such as alternative security organisations or other institutional choices in autocracies, or more theoretically hone in on specific predictors with the aim of establishing causal relationships.   

But additionally, given how scarcely secret police are formed and how enduring they are, this research also highlights their internal dynamics as a subject for future study. The recent wave of studies focusing on particular secret police cases in, for instance, Eastern Germany \citep{Steinert_2023}, Poland \citep{hager2022does,Thomson_2020,thomson2024bureaucratic}, Taiwan \citep{Liu_Su_Wang_2026}, and Argentina \citep{Scharpf_Gläßel_2020} is helpful in this regard. But future work should supplement this with a greater focus on comparison and, eventually, extending global data on secret police existence with more information regarding their activities, size, and surveillance methods.

\appendix
\newpage

\maketitle

\setcounter{figure}{0}
\setcounter{table}{0}
\setcounter{footnote}{0}
\renewcommand\thefigure{A.\arabic{figure}}
\renewcommand\thetable{A.\arabic{table}}
\renewcommand\theHfigure{A.\arabic{figure}}
\renewcommand\theHtable{A.\arabic{table}}

\section{Data Sources and Software}

Table \ref{tab:datasources} lists the data sources for all independent variables included in the models presented here.
As discussed in the main paper, the data on secret police existence used to construct the dependent variable 'secret police formation' comes from \citet*{choulis2024preventing}. In accessing some of the data sources listed in Table \ref{tab:datasources}, we benefited from \texttt{Peacesciencer} \citep{miller2022peacesciencer} and the WEP Dataverse \citep{graham2019international}. 

We utilised the packages \texttt{cshapes} \citep{schvitz2022mapping} and \texttt{sf} \citep{sf} for spatial processing.
Multiple imputation of the missing values was facilitated by \texttt{Amelia} \citep{honaker2011}, 
while \texttt{glmnet} was used for estimating Lasso regression and cross-validation. 
The \texttt{margins} \citet{margins} package was used to calculate Average Marginal Effects (AME). 

\begin{table}[t]
\centering
  \begin{threeparttable}
    \caption{Variables and data sources}    
    \label{tab:datasources}
    \scriptsize
\begin{tabular}{l l p{0.3cm} l l}         
\toprule
\textbf{Variable} & \textbf{Data source} & & \textbf{Variable} & \textbf{Data source} \\
\midrule
Intrastate conflict: Dummy & \cite{davies2024organized} & & Clientelism & \cite{Coppedge_Gerring_Knutsen_Lindberg_Teorell_etal._2023} \\
Intrastate conflict: Years since & \cite{davies2024organized} & & Ethnically excluded pop. & \cite{Vogt_Bormann_Rüegger_Cederman_Hunziker_Girardin_2015} \\
Neighborhood intrastate conflicts & \cite{davies2024organized} & & Counterbalancing & \cite{Pilster_Bohmelt_2011} \\
Democracy score & \cite{Marshall_Gurr_2020} & & Military Exp. & \cite{Barnum_Fariss_Markowitz_Morales_2024} \\ 
Coup attempt: Dummy & \cite{Powell_Thyne_2011} & & Oil production & \cite{Ross2015_oildata} \\  
Coup attempt: Years since & \cite{Powell_Thyne_2011} & & Gas production & \cite{Ross2015_oildata} \\  
Human rights & \cite{Fariss_2019} & & Election year & \cite{Hyde_Marinov_2012} \\  
State capacity & \cite{Hanson_Sigman_2021} & & Leader duration & \cite{Mattes_Leeds_Matsumura_2016} \\  
Bureaucratic capacity & \cite{Coppedge_Gerring_Knutsen_Lindberg_Teorell_etal._2023} & & Regime duration & \cite{Mattes_Leeds_Matsumura_2016} \\ 
Fiscal capacity & \cite{Coppedge_Gerring_Knutsen_Lindberg_Teorell_etal._2023} & & Neighborhood unsucc. coups & \cite{Powell_Thyne_2011} \\ 
Territorial control & \cite{Coppedge_Gerring_Knutsen_Lindberg_Teorell_etal._2023} & & Neighborhood succ. coups & \cite{Powell_Thyne_2011} \\  
CSO entry and exit & \cite{Coppedge_Gerring_Knutsen_Lindberg_Teorell_etal._2023} & & Int. rivalry: Dummy & \cite{Thompson_Dreyer_2011} \\  
CSO repression & \cite{Coppedge_Gerring_Knutsen_Lindberg_Teorell_etal._2023} & & Int. rivalry: Count & \cite{Thompson_Dreyer_2011} \\  
CSO participatory env. & \cite{Coppedge_Gerring_Knutsen_Lindberg_Teorell_etal._2023} & & MID: Dummy & \cite{Palmer_etal_2022} \\
CSO anti-system mov. & \cite{Coppedge_Gerring_Knutsen_Lindberg_Teorell_etal._2023} & & MID: Years since & \cite{Palmer_etal_2022} \\
CSO strength & \cite{Coppedge_Gerring_Knutsen_Lindberg_Teorell_etal._2023} & & MID: Count & \cite{Palmer_etal_2022} \\    
Regime change & \cite{Mattes_Leeds_Matsumura_2016} & & Protest & \cite{Chenoweth_D’Orazio_Wright_2014} \\   
Urban population & \cite{WorldBank_2021} & & Neighborhood protest & \cite{Chenoweth_D’Orazio_Wright_2014} \\  
Economic growth & \cite{WorldBank_2021} & & GDP p.c. & \cite{WorldBank_2021} \\ 
Personalisation & \cite{Geddes_Wright_Frantz_2018} & & Population & \cite{WorldBank_2021} \\  
Ideological legitimisation & \cite{Coppedge_Gerring_Knutsen_Lindberg_Teorell_etal._2023} & & Affiliated forces & \cite{debruin2021mapping} \\
Neighborhood: Secret Police & \cite{choulis2024preventing} & & Counterweight & \cite{debruin2021mapping} \\
\bottomrule
\end{tabular}
  \end{threeparttable}
\end{table}

\hide{\begin{table}[t]
\centering
  \begin{threeparttable}
    \caption{Variables and data sources}    
    \label{tab:datasources}
    \scriptsize
\begin{tabular}{l l}         \toprule
        \textbf{Variable} & \textbf{Data source} \\
        \midrule
  Intrastate conflict: Dummy (UCDP) &  \cite{davies2024organized}\\
  Intrastate conflict: Years since (UCDP) &  \cite{davies2024organized}\\
  Intrastate conflict: Neighbour (UCDP) &  \cite{davies2024organized}\\
  Democracy score (Polity5)  &  \cite{Marshall_Gurr_2020} \\ 
  Coup attempt: Dummy  & \cite{Powell_Thyne_2011} \\  
  Coup attempt: Years since & \cite{Powell_Thyne_2011} \\  
  Human rights (Latent score) & \cite{Fariss_2019} \\  
  State capacity (Latent score)  & \cite{Hanson_Sigman_2021} \\  
  Bureaucratic capacity (V-DEM) & \cite{Coppedge_Gerring_Knutsen_Lindberg_Teorell_etal._2023} \\ 
  Fiscal capacity (V-DEM) & \cite{Coppedge_Gerring_Knutsen_Lindberg_Teorell_etal._2023} \\ 
  Territorial control (V-DEM) & \cite{Coppedge_Gerring_Knutsen_Lindberg_Teorell_etal._2023} \\  
 CSO entry and exit (V-DEM) & \cite{Coppedge_Gerring_Knutsen_Lindberg_Teorell_etal._2023} \\  
 CSO repression (V-DEM) & \cite{Coppedge_Gerring_Knutsen_Lindberg_Teorell_etal._2023}  \\  
 CSO participatory environment (V-DEM) &  \cite{Coppedge_Gerring_Knutsen_Lindberg_Teorell_etal._2023} \\
 CSO anti-system movements (V-DEM) & \cite{Coppedge_Gerring_Knutsen_Lindberg_Teorell_etal._2023} \\
 CSO strength  (V-DEM)  & \cite{Coppedge_Gerring_Knutsen_Lindberg_Teorell_etal._2023} \\    
 Regime change (CHISOLS) & \cite{Mattes_Leeds_Matsumura_2016} \\   
 Urban population (\%) & \cite{WorldBank_2021} \\  
 Economic growth & \cite{WorldBank_2021} \\ 
 Personalisation (Latent score)  &  \cite{Geddes_Wright_Frantz_2018} \\  
 Ideological legitimisation (V-Dem) & \cite{Coppedge_Gerring_Knutsen_Lindberg_Teorell_etal._2023} \\
 Clientelism (V-Dem) & \cite{Coppedge_Gerring_Knutsen_Lindberg_Teorell_etal._2023} \\
 Ethnically excluded population (\%) & \cite{Vogt_Bormann_Rüegger_Cederman_Hunziker_Girardin_2015} \\
 Counterbalancing & \cite{Pilster_Bohmelt_2011} \\
 Military expenditures (Latent score, \emph{ln}) & \cite{Barnum_Fariss_Markowitz_Morales_2024} \\
 Oil production (Financial value, \emph{ln}) & \cite{Ross2015_oildata} \\
 Gas production (Financial value, \emph{ln}) &  \cite{Ross2015_oildata} \\
 Election year (NELDA) & \cite{Hyde_Marinov_2012} \\
 Leader duration (CHISOLS) & \cite{Mattes_Leeds_Matsumura_2016} \\
 Regime duration (CHISOLS) & \cite{Mattes_Leeds_Matsumura_2016} \\
 Successful coups in neighbourhood & \cite{Powell_Thyne_2011} \\
 Unsuccessful coups in neighbourhood & \cite{Powell_Thyne_2011} \\
 International rivalry: Dummy & \cite{Thompson_Dreyer_2011} \\
 International rivalry: Count & \cite{Thompson_Dreyer_2011} \\
 MID: Dummy &  \cite{Palmer_etal_2022} \\
 MID: Years since &  \cite{Palmer_etal_2022} \\
 MID: Count & \cite{Palmer_etal_2022}\\
 Protest (Latent score) & \cite{Chenoweth_D’Orazio_Wright_2014} \\
 Neighbour protest (Latent score) & \cite{Chenoweth_D’Orazio_Wright_2014} \\
 GDP p.c. (Logged) & \cite{WorldBank_2021}  \\
 Population (Logged) &  \cite{WorldBank_2021} \\
 affiliated & \cite{debruin2021mapping} \\
 counterweight & \cite{debruin2021mapping} \\
   
    \bottomrule
    \end{tabular}
  \end{threeparttable}
\end{table}}

\section{Main Results}

Table \ref{tab:modeloutput_main} presents the full results for the four models underlying the marginal effects illustrated in Figure 1 of the main paper.

\begin{table}[!tbp]
\caption{Full results table underlying figure 1. DV: Secret Police. Standard errors are reported in parentheses below the average marginal effects.\label{tab:modeloutput_main}} 
{\centering
\begin{tabular}{lrrrrr}
\toprule
\multicolumn{1}{l}{\bfseries \textbf{Variable}}&\multicolumn{2}{c}{\bfseries LASSO}&\multicolumn{1}{c}{\bfseries }&\multicolumn{2}{c}{\bfseries Stepwise}\tabularnewline
\cline{2-3} \cline{5-6}
\multicolumn{1}{l}{}&\multicolumn{1}{c}{Logit}&\multicolumn{1}{c}{Cloglog}&\multicolumn{1}{c}{}&\multicolumn{1}{c}{Logit}&\multicolumn{1}{c}{Cloglog}\tabularnewline
\midrule
Intercept&-8.7775&-8.4632&&-11.2836&-10.9846\tabularnewline
&(1.2660)&(1.2162)&&(1.4989)&(1.4448)\tabularnewline
Leader Duration&-0.0543&-0.0506&&\textemdash&\textemdash\tabularnewline
&(0.0428)&(0.0420)&&&\tabularnewline
GDP p.c.&0.1813&0.1722&&0.2302&0.2159\tabularnewline
&(0.1170)&(0.1149)&&(0.1147)&(0.1113)\tabularnewline
Military Expenditures&0.3192&0.2854&&0.6931&0.6618\tabularnewline
&(0.1505)&(0.1474)&&(0.1764)&(0.1714)\tabularnewline
MID: Count&-0.2471&-0.2342&&-0.2583&-0.2449\tabularnewline
&(0.1535)&(0.1486)&&(0.1451)&(0.1412)\tabularnewline
Protest&0.4473&0.4289&&0.4581&0.4479\tabularnewline
&(0.2821)&(0.2783)&&(0.2681)&(0.2629)\tabularnewline
Regime Duration&-0.0408&-0.0381&&-0.0671&-0.0649\tabularnewline
&(0.0234)&(0.0229)&&(0.0205)&(0.0199)\tabularnewline
Internat. Rivalry: Count&0.1149&0.1141&&\textemdash&\textemdash\tabularnewline
&(0.0741)&(0.0716)&&&\tabularnewline
Neighborhood: Successful Coups&0.5795&0.5554&&0.6280&0.5982\tabularnewline
&(0.3751)&(0.3671)&&(0.3558)&(0.3429)\tabularnewline
Human Rights&-0.0550&-0.0667&&\textemdash&\textemdash\tabularnewline
&(0.2500)&(0.2439)&&&\tabularnewline
CSO Anti-system Movements&0.3416&0.3233&&0.4115&0.3993\tabularnewline
&(0.1837)&(0.1797)&&(0.1722)&(0.1683)\tabularnewline
CSO Participatory Environment&-0.1142&-0.1267&&\textemdash&\textemdash\tabularnewline
&(0.2149)&(0.2123)&&&\tabularnewline
CSO Repression&-0.3735&-0.3257&&-0.8024&-0.7668\tabularnewline
&(0.2352)&(0.2274)&&(0.2071)&(0.1974)\tabularnewline
Personalisation&1.6071&1.4582&&1.6504&1.5807\tabularnewline
&(0.7918)&(0.7691)&&(0.7176)&(0.6972)\tabularnewline
\bottomrule
\end{tabular}}
\end{table}

\hide{
\begin{table}[h!]
\centering
\caption{Full results table underlying Figure \ref{fig:marginal_effects_cw} -- DV: Counterweight Paramilitary. Standard errors are reported in parentheses below the average marginal effects. Coefficients for covariates that were not selected by the Lasso model are represented by the symbol \textemdash. }
\label{tab:modeloutput} 
\renewcommand{\arraystretch}{1.0}
\begin{tabular}[t]{l r r r r r}
\toprule
& \multicolumn{2}{c}{\textbf{LASSO}} & & \multicolumn{2}{c}{\textbf{Stepwise}} \\ \cline{2-3} \cline{5-6} \rule{-3pt}{14pt}
\textbf{Variable}   & Logit & Cloglog & & Logit & Cloglog \\
\midrule
Intercept                               & $-$7.7292 & $-$7.6051 & & $-$10.8003   & $-$10.5246 \\
                                        & (1.1229)  & (1.0976)  & & (1.4622)    & (1.4048) \\[1ex]
Gas production                          & \multicolumn{1}{c}{~~\textemdash}    & \multicolumn{1}{c}{~~\textemdash}    & & $-$0.0574   & $-$0.0569\\
                                        &   &   & & (0.0267)    & (0.0262)\\[1ex]
GDP p.c.                                & 0.1264    & 0.1182    & & 0.2172      & 0.2055 \\
                                        & (0.1113)  & (0.1092)  & & (0.1149)    & (0.1115) \\[1ex]
Military Expenditures                   & 0.2504    & 0.2395    & & 0.6268      & 0.5965 \\ 
                                        & (0.1403)  & (0.1383)  & & (0.1770)    & (0.1721) \\[1ex]
Militarized Interstate Dispute: Count   & \multicolumn{1}{c}{~~\textemdash} & \multicolumn{1}{c}{~~\textemdash} & & $-$0.2561   & $-$0.2411 \\
                                        &  &   & & (0.1493)    & (0.1454) \\[1ex]
Protest                                 & 0.4080    & 0.4036    & & 0.4298      & 0.4252 \\
                                        & (0.2557)  &
                        (0.2512)  & & (0.2741)    & (0.2678) \\[1ex]
Regime Duration                         & $-$0.0567 & $-$0.0556 & & $-$0.0653   & $-$0.0635\\
                                        & (0.0209)  & (0.0205)  & & (0.0205)    & (0.0199) \\[1ex]
International Rivalry: Count            & 0.1137    & 0.1072    & & 0.2042      & 0.1988 \\
                                        & (0.1590)  & (0.1549)  & & (0.1626)    & (0.1566) \\[1ex]
International Rivalry: Dummy            & 0.7871    & 0.7942    & & 0.7305      & 0.7442\\
                                        & (0.4265)  & (0.4186)  & & (0.4427)    & (0.4322) \\[1ex]
Human Rights                            & 0.0190 & 0.0176 & & \multicolumn{1}{c}{~~\textemdash}      &  \multicolumn{1}{c}{~~\textemdash} \\
                                        & (0.2492)  & (0.2453)  & &     &  \\[1ex]
CSO anti-system movements               & 0.4188    & 0.4161    & & 0.4191      & 0.4100 \\
                                        & (0.1729)  & (0.1694)  & & (0.1705)    & (0.1663) \\[1ex]
CSO participatory environment           & $-$0.1128 & $-$0.1184   & &  \multicolumn{1}{c}{~~\textemdash}    & \multicolumn{1}{c}{~~\textemdash} \\
                                        & (0.2124)  & (0.2099)  & &     & \\[1ex]
CSO repression                          & $-$0.4281 & $-$0.4147 & & $-$0.7532   & $-$0.7258 \\
                                        & (0.2360)  & (0.2321)  & & (0.2120)    & (0.2020) \\[1ex]

Ideological Legitimisation          & \multicolumn{1}{c}{~~\textemdash}  & \multicolumn{1}{c}{~~\textemdash}  & &  $-$0.4976    & $-$0.4833 \\
                       &  &   & & (0.2179)    &(0.2129)  \\ [1ex]

Personalisation                         & \multicolumn{1}{c}{~~\textemdash}    & \multicolumn{1}{c}{~~\textemdash}    & & 1.6876      & 1.6378 \\
                        &  &   & & (0.7282)    & (0.7051) \\ [1ex]
\hline
\bottomrule
\end{tabular}
\end{table}
}

\section{Results for Counterweight Dependent Variable}

\begin{table}[!tbp]
\caption{Full results table underlying figure A.1. DV: Counterweight Paramilitary. Standard errors are reported in parentheses below the average marginal effects.\label{tab:modeloutput_cw}} 
{\centering
\begin{tabular}{lrrrrr}
\toprule
\multicolumn{1}{l}{\bfseries \textbf{Variable}}&\multicolumn{2}{c}{\bfseries LASSO}&\multicolumn{1}{c}{\bfseries }&\multicolumn{2}{c}{\bfseries Stepwise}\tabularnewline
\cline{2-3} \cline{5-6}
\multicolumn{1}{l}{}&\multicolumn{1}{c}{Logit}&\multicolumn{1}{c}{Cloglog}&\multicolumn{1}{c}{}&\multicolumn{1}{c}{Logit}&\multicolumn{1}{c}{Cloglog}\tabularnewline
\midrule
Intercept&1.2403&0.4278&&1.0404&0.3259\tabularnewline
&(0.2914)&(0.1781)&&(0.3071)&(0.1869)\tabularnewline
Leader Duration&0.0095&0.0060&&\textemdash&\textemdash\tabularnewline
&(0.0067)&(0.0043)&&&\tabularnewline
GDP p.c.&-0.0471&-0.0293&&-0.0136&-0.0095\tabularnewline
&(0.0328)&(0.0217)&&(0.0384)&(0.0249)\tabularnewline
Military Expenditures&-0.0382&-0.0259&&-0.0830&-0.0593\tabularnewline
&(0.0446)&(0.0279)&&(0.0400)&(0.0255)\tabularnewline
MID: Count&0.0267&0.0146&&0.0110&0.0052\tabularnewline
&(0.0297)&(0.0186)&&(0.0312)&(0.0207)\tabularnewline
Protest&-0.1064&-0.0619&&-0.1677&-0.0942\tabularnewline
&(0.0705)&(0.0439)&&(0.0610)&(0.0386)\tabularnewline
Regime Duration&-0.0065&-0.0038&&-0.0064&-0.0041\tabularnewline
&(0.0043)&(0.0029)&&(0.0037)&(0.0025)\tabularnewline
Internat. Rivalry: Count&-0.0096&-0.0060&&\textemdash&\textemdash\tabularnewline
&(0.0243)&(0.0155)&&&\tabularnewline
Neighborhood: Successful Coups&-0.0247&-0.0269&&0.0134&-0.0053\tabularnewline
&(0.1180)&(0.0718)&&(0.1156)&(0.0706)\tabularnewline
Human Rights&0.0572&0.0264&&\textemdash&\textemdash\tabularnewline
&(0.0568)&(0.0346)&&&\tabularnewline
CSO Anti-system Movements&0.1850&0.1183&&0.1553&0.0998\tabularnewline
&(0.0506)&(0.0317)&&(0.0404)&(0.0250)\tabularnewline
CSO Participatory Environment&-0.2478&-0.1602&&\textemdash&\textemdash\tabularnewline
&(0.0721)&(0.0426)&&&\tabularnewline
CSO Repression&0.0983&0.0637&&0.0466&0.0262\tabularnewline
&(0.0563)&(0.0348)&&(0.0428)&(0.0265)\tabularnewline
Personalisation&-0.0707&-0.0406&&0.1209&0.1076\tabularnewline
&(0.2502)&(0.1530)&&(0.2351)&(0.1482)\tabularnewline
\bottomrule
\end{tabular}}
\end{table}

\begin{figure}
    \centering
    \includegraphics[width=\linewidth]{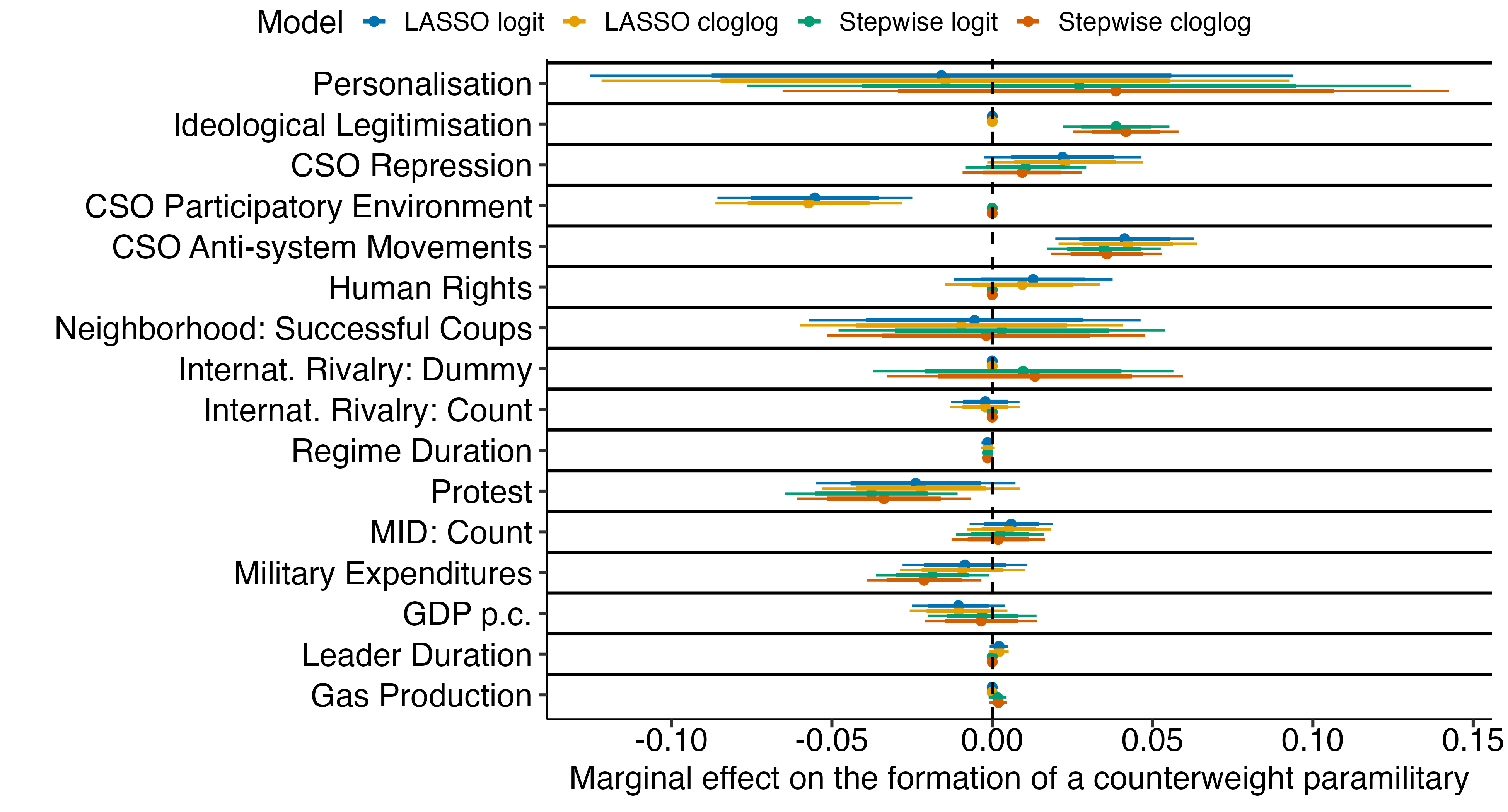}
    \caption{Marginal effects for model with counterweight as target variable}
    \label{fig:marginal_effects_cw}
\end{figure}

\hide{
\begin{table}[!tbp]
\caption{Full results table underlying figure 1. Standard errors are reported in parentheses below the average marginal effects.\label{tab:modeloutput}} 
{\centering
\begin{tabular}{lrrrrr}
\toprule
\multicolumn{1}{l}{\bfseries \textbf{Variable}}&\multicolumn{2}{c}{\bfseries LASSO}&\multicolumn{1}{c}{\bfseries }&\multicolumn{2}{c}{\bfseries Stepwise}\tabularnewline
\cline{2-3} \cline{5-6}
\multicolumn{1}{l}{}&\multicolumn{1}{c}{Logit}&\multicolumn{1}{c}{Cloglog}&\multicolumn{1}{c}{}&\multicolumn{1}{c}{Logit}&\multicolumn{1}{c}{Cloglog}\tabularnewline
\midrule
Intercept&1.2459&0.4118&&1.6777&0.5215\tabularnewline
&(0.3551)&(0.2044)&&(0.3303)&(0.2223)\tabularnewline
Leader Duration&0.0111&0.0066&&\textemdash&\textemdash\tabularnewline
&(0.0061)&(0.0038)&&&\tabularnewline
GDP p.c.&-0.0362&-0.0206&&-0.1047&-0.0261\tabularnewline
&(0.0291)&(0.0179)&&(0.0308)&(0.0180)\tabularnewline
Military Expenditures&-0.0493&-0.0340&&-0.1092&-0.0620\tabularnewline
&(0.0449)&(0.0288)&&(0.0406)&(0.0288)\tabularnewline
Protest&-0.1032&-0.0599&&-0.1419&-0.0707\tabularnewline
&(0.0723)&(0.0444)&&(0.0785)&(0.0458)\tabularnewline
Regime Duration&-0.0053&-0.0028&&-0.0048&-0.0015\tabularnewline
&(0.0033)&(0.0022)&&(0.0030)&(0.0018)\tabularnewline
International Rivalry: Count&0.0824&0.0527&&0.0838&0.0453\tabularnewline
&(0.0654)&(0.0404)&&(0.0792)&(0.0478)\tabularnewline
International Rivalry: Dummy&-0.2630&-0.1622&&-0.2727&-0.1576\tabularnewline
&(0.1177)&(0.0719)&&(0.1171)&(0.0718)\tabularnewline
Successful Coups in Neighbors&-0.0591&-0.0474&&\textemdash&\textemdash\tabularnewline
&(0.1150)&(0.0711)&&&\tabularnewline
Human Rights&0.0653&0.0304&&\textemdash&\textemdash\tabularnewline
&(0.0545)&(0.0342)&&&\tabularnewline
CSO Anti-system Movements&0.1718&0.1108&&0.1845&0.1094\tabularnewline
&(0.0429)&(0.0268)&&(0.0442)&(0.0256)\tabularnewline
CSO Participatory Environment&-0.2510&-0.1592&&-0.2494&-0.1514\tabularnewline
&(0.0533)&(0.0341)&&(0.0531)&(0.0352)\tabularnewline
CSO Repression&0.1275&0.0811&&0.1922&0.1238\tabularnewline
&(0.0492)&(0.0317)&&(0.0478)&(0.0293)\tabularnewline
Personalisation&-0.0282&0.0012&&0.0127&0.0731\tabularnewline
&(0.2213)&(0.1382)&&(0.2091)&(0.1315)\tabularnewline
\bottomrule
\end{tabular}}
\end{table}
}
In Table \ref{tab:modeloutput_cw}, we present the results of four models where the models from Table \ref{tab:modeloutput_main} are replicated using the formation of a counterweight paramilitary \citep{debruin2021mapping} as dependent variable. 
These results, visualised in Figure \ref{fig:marginal_effects_cw}, allow us to asses to what extent the main results hold across different types of state security organisations or are specific to secret police. These results indicate that while several of the covariates found to be important in the main model are also statistically significant here, most of these actually exhibit \emph{opposing} effect directions. 
For counterweight paramilitaries, our models thus uncover negative effects for protest, GDP per capita, and military spending, but a positive effect for civil society repression. The existence of anti-system civil society organizations is again found to be positive and statistically significant, while regime duration again exhibits a negative and statistically significant effect. 
In contrast to secret police, counterweights appear unaffected by personalism, successful coups in the neighbourhood, and the existence of international rivalries. These results can be taken to suggest that secret police and counterweights are created when the ruler faces different threats, further supporting their analytical distinction.

\section{Additional results}

\begin{figure}
    \centering
    \includegraphics[width=\linewidth]{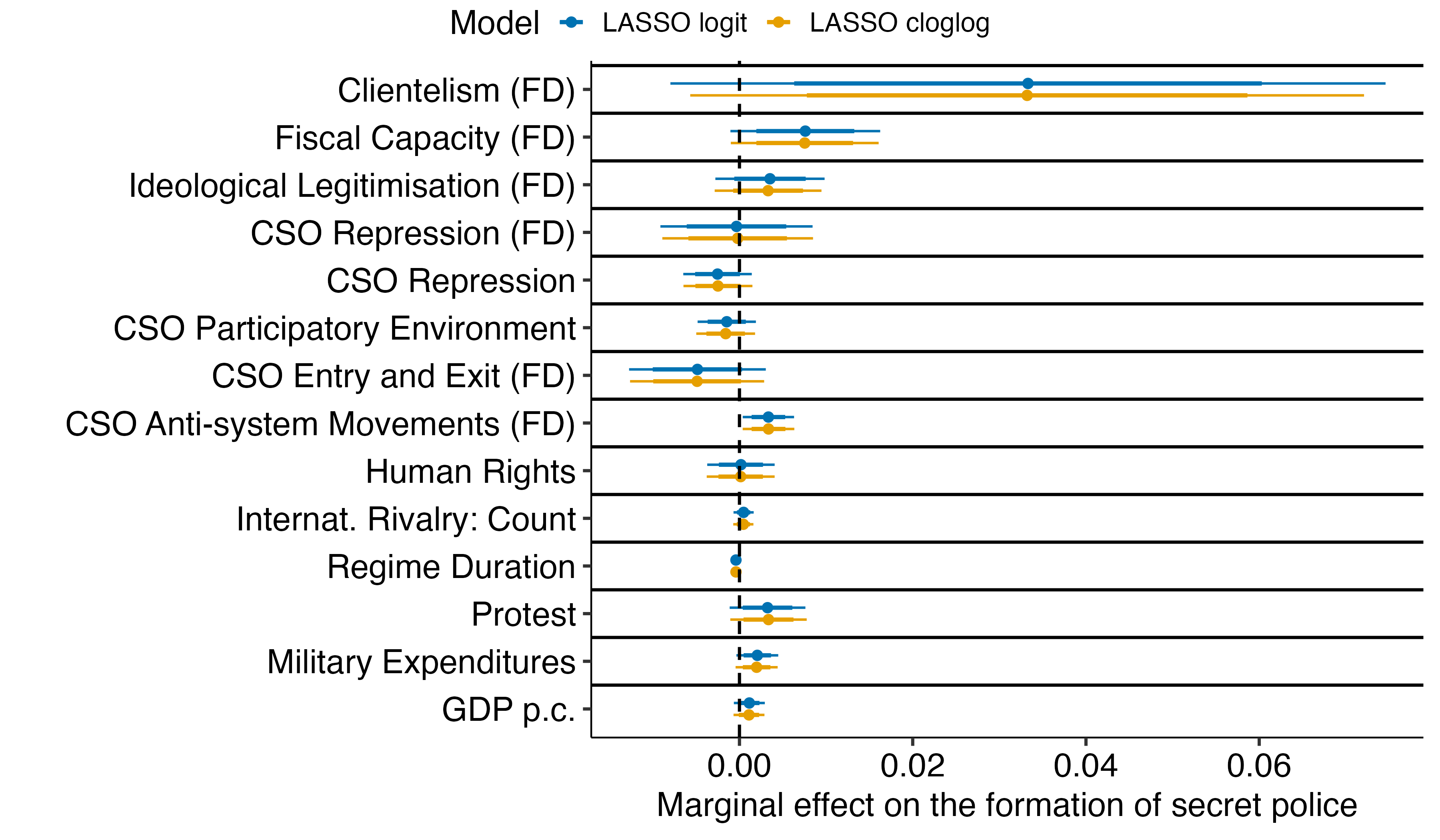}
    \caption{Marginal effects for model with first differences as covariates.}
    \label{fig:marginal_effects_diff}
\end{figure}

In Table \ref{tab:modeloutput_diff} and Figure \ref{fig:marginal_effects_diff}, we show the results of additional models that, in addition to the variables presented in Table 2 in the paper, also include the first differences for all continuous variables included there and the time since the last realisation for all binary variables. There, we focus exclusively on the Lasso, as these models are comparatively better at handling the sheer number of covariates we now have to select from.

As may be expected, the results visible in Figure \ref{fig:marginal_effects_diff} look somewhat different than those of the main models, now also including first differences for some covariates that had previously not been selected. 
In particular, these results point to the role of clientelism and fiscal state capacity \emph{increasing} the probability of secret police formation. 
However, there are also strong similarities between the results, even though the models visualized here included almost twice the number of covariates featured in the main models: structural, regime-external threats such as protests and anti-system mobilisation are again found to increase the probability of secret police formation, as are indicators of the regime's ability to finance such costly security actors. The additional result that positive changes in fiscal state capacity increase the probability of secret police being established aligns with this latter group of findings. 
It may be possible that increases in clientelism are linked to a ruler's efforts of establishing personalised control over the regime. 

\begin{table}[!tbp]
\caption{Full results table underlying figure A.2. DV: Secret Police. Standard errors are reported in parentheses below the average marginal effects.\label{tab:modeloutput_diff}} 
{\centering
\begin{tabular}{lrr}
\toprule
\multicolumn{1}{l}{\textbf{Variable}}&\multicolumn{1}{c}{Logit}&\multicolumn{1}{c}{Cloglog}\tabularnewline
\midrule
Intercept&-7.6030&-7.4443\tabularnewline
&(1.1313)&(1.1002)\tabularnewline
GDP p.c.&0.1431&0.1348\tabularnewline
&(0.1125)&(0.1097)\tabularnewline
Military Expenditures&0.2587&0.2427\tabularnewline
&(0.1496)&(0.1468)\tabularnewline
Protest&0.4069&0.4107\tabularnewline
&(0.2731)&(0.2682)\tabularnewline
Regime Duration&-0.0510&-0.0496\tabularnewline
&(0.0218)&(0.0214)\tabularnewline
Internat. Rivalry: Count&0.0599&0.0548\tabularnewline
&(0.0740)&(0.0723)\tabularnewline
Human Rights&0.0214&0.0172\tabularnewline
&(0.2492)&(0.2451)\tabularnewline
CSO Anti-system Movements (FD)&0.4194&0.4110\tabularnewline
&(0.1783)&(0.1743)\tabularnewline
CSO Entry and Exit (FD)&-0.6089&-0.5991\tabularnewline
&(0.4970)&(0.4743)\tabularnewline
CSO Participatory Environment&-0.1842&-0.1951\tabularnewline
&(0.2137)&(0.2101)\tabularnewline
CSO Repression&-0.3177&-0.3043\tabularnewline
&(0.2489)&(0.2443)\tabularnewline
CSO Repression (FD)&-0.0430&-0.0252\tabularnewline
&(0.5626)&(0.5438)\tabularnewline
Ideological Legitimisation (FD)&0.4423&0.4048\tabularnewline
&(0.3990)&(0.3804)\tabularnewline
Fiscal Capacity (FD)&0.9547&0.9234\tabularnewline
&(0.5355)&(0.5134)\tabularnewline
Clientelism (FD)&4.1815&4.0689\tabularnewline
&(2.5907)&(2.3747)\tabularnewline
\bottomrule
\end{tabular}}
\end{table}

\newpage
\bibliographystyle{chicago}
\bibliography{library.bib}

\end{document}